\documentstyle[psfig]{mn}

\newif\ifAMStwofonts

\newcommand{\Rsolar}{\mbox{$R_{\odot}$}}
\newcommand{\Msolar}{\mbox{$M_{\odot}$}}

\title[KPD\,1930+2752]
{KPD\,1930+2752 -- a candidate Type Ia supernova progenitor.}
\author[P. F. L. Maxted, T. R. Marsh \& R. C. North]
       {P. F. L. Maxted, T. R. Marsh \& R. C. North \\
        University of Southampton, Department of Physics \& Astronomy,
        Highfield, Southampton, S017 1BJ, UK}
\date{Accepted ---
      Received ---}

\pagerange{\pageref{firstpage}--\pageref{lastpage}}
\pubyear{1998}

\begin{document}

\maketitle

\label{firstpage}

\begin{abstract}
 We present spectra of the pulsating sdB star KPD\,1930+2752 which confirm
that this star is a binary. The radial velocities measured from the H$\alpha$
 and HeI\,6678\AA\ spectral lines vary sinusoidally with the same period
($2^{\rm h} 17^{\rm m}$) as the ellipsoidal variability seen by
Bill\`{e}res et~al. (2000). The amplitude of the orbital motion
(349.3$\pm$2.7\,km\,s$^{-1}$) combined with the canonical mass for sdB stars
(0.5\Msolar) implies a total mass for the binary of 1.47$\pm$0.01\Msolar. The
unseen companion star is almost certainly a white dwarf star. The binary will
merge within $\sim$ 200 million years due to gravitational wave radiation. The
accretion of helium and other elements heavier than hydrogen onto the white
dwarf which then exceeds the Chandrasekhar mass (1.4\Msolar) is a viable model
for the cause of Type~Ia supernovae. KPD\,1930+2752 is the first star to be
discovered which is a good candidate for the progenitor of a Type~Ia supernova
of this type which will merge on an astrophysically interesting timescale.
\end{abstract} 

\section{Introduction}

 Type~Ia supernovae (SNe~Ia) are one of the most important tools for
observational cosmology because there appears to be a relatively small spread
in their peak optical brightness around $M_{\rm V} = -19.6$ and  they can be
seen out to cosmological distances (z$\sim$1) so they can be used to measure
cosmological parameters, e.g., the cosmological constant, $\Lambda$ (Riess
et~al. 1998, Perlmutter et~al. 1999). However, the peak optical brightnesses
of  SNe~Ia are not uniform, they are correlated with the shape of the
lightcurve and vary by about one magnitude. Meaningful measurements of
cosmological parameters require this variation to be calibrated, e.g., the
non-zero value of $\Lambda$ measured by Perlmutter et~al. is required by
supernovae at z$\sim$0.5 being about 0.3 magnitudes too bright compared to a
non-accelerating ($\Lambda$=0) Universe. The corrections to peak brightnesses
have to be empirical because it is still not yet clear what causes SNe~Ia.  

 Type~Ia supernovae near maximum light show no hydrogen or helium lines but do
show strong silicon lines. The absence in the spectrum of the two most common
elements in the Universe dramatically reduces the number of potential
progenitors, as does their appearance in old stellar populations, e.g.,
elliptical galaxies. All the most likely models for progenitors feature an
accreting white dwarf (Leibundgut 2000; Branch et~al. 1995) which ignites
carbon in its core either because it has reached the Chandrasekhar mass
(1.4\Msolar) or because ignition of accumulated helium causes compression of
the core and a so-called ``edge-lit detonation''. This explains the fast rise
times for SNe~Ia,  the lack of hydrogen and helium and the fairly uniform peak
brightness. To initiate the explosion, the white dwarf must accrete material
from a companion star. Two models for the companion star which  have gained
popularity in recent times are  super-soft sources and double degenerates. 

Super-soft sources are white dwarfs which accrete hydrogen from a
non-degenerate star at a rate just sufficient to support steady nuclear
burning on the surface of the white dwarf (Kahabka \& van den Heuvel 1997).
This leads to an accumulation of material on the white dwarf but it is not
clear whether or not the accretion rate stays within the required range 
sufficiently long for a SNe~Ia explosion to result or that there are a
sufficient number of these binaries to explain the observed rate of SNe~Ia.
Neither is it clear that these binaries have a sufficiently long lifetime to
be observed in elliptical galaxies (Yungelson \& Livio 2000; Yungelson et~al.
1995). 

The double degenerate model posits two white dwarfs with an orbital period of
a few hours which merge due to the loss of gravitational wave radiation.  Two
drawbacks with this model have been a poor understanding of how such a
detonation might be initiated and the lack of observed progenitors. There are
now many double degenerates known (Marsh, Dhillon \& Duck 1995; Moran, Marsh
\& Bragaglia, 1997; Moran, Maxted \& Marsh 2000) but none have both a
sufficiently short orbital period and a total mass in excess of 1.4\Msolar.
However,  at least one good candidate for the progenitor of an edge-lit
detonation is known (WD\,1704+481.2; Maxted et~al. 2000).

 KPD\,1930+2752 was identified as a subdwarf-B (sdB) star in the Kitt Peak --
Downes survey of UV excess objects near the Galactic plane (Downes 1986).
Photometry in the Cousins {\it BVRI}\ system by Allard et~al. (1994) and
Str\"{o}mgren photometry by Wesemael et~al. (1992)  revealed nothing
exceptional about this star other than that it has a low reddening and that
there is no evidence for companion to this star. Of the 100 subdwarfs in the
study of Allard et~al., 31 show evidence for a companion which, when some
estimate of the selection effects is made, suggests that more than half of
these stars are binaries.

 The binary fraction of sdB stars is a matter of some interest because the
properties of these stars suggest they have lost a substantial fraction of
their mass, perhaps due to interactions with a companion star (Heber~1986;
Saffer~et~al. 1994; Iben \& Livio 1993). In addition to the  composite
spectrum binaries identified by Allard et~al., several binary sdB stars have
been identified from the Doppler shift of the spectral lines due to the orbital
motion (Saffer, Livio \& Yungelson 1998). This technique has the advantage of
being sensitive to the companion star whatever its type and can detect sdB
stars with white dwarf companions which would be missed by almost any
photometric technique.  Direct evidence for white dwarf companions to sdB
stars is seen in the eclipsing sdB\,--\,white dwarf binary KPD\,0422+5421
(Orosz \& Wade 2000). 

 Other clues to the properties of sdB stars comes from the EC~14026 stars --
sdB stars showing {\it p}-mode pulsations with periods of a few minutes (Koen
et~al. 1998; Fontaine et~al. 1998; O'Donoghue et~al. 1999). High speed
photometry by Bill\`{e}res et~al. (2000) identified KPD\,1930+2752 as the
14th EC~14026 star known. The photometry also showed variability with a period
of $2^{\rm h}17{\rm m}$ with an amplitude of 1.4\,percent. This is much longer
than the pulsation periods of EC~14026 stars. The lightcurve folded on this
period shows an almost sinusoidal variation with two minima per cycle, but
with one minimum being slightly deeper than the other. This was interpreted as
being the ellipsoidal variation due to the rotation of a star distorted by the
presence of a companion star.

 In this paper we present spectroscopy of the the H$\alpha$ and HeI\,6678\AA\
spectral lines which confirms that KPD\,1930+2752 is a binary star with an
orbital period of $2^{\rm h}17^{\rm m}$. We also show that the total mass of
the binary exceeds the Chandrasekhar mass and conclude that KPD\,1930+2752 is
the first good candidate Type~Ia supernova progenitor which will explode due
to the accretion of helium and other elements heavier than hydrogen onto a
white dwarf on an astrophysically interesting timescale. 

\section{Observations and reductions}
 Observations were obtain with the Isaac Newton Telescope at the Observatario
Roque de los Muchachos on the Island of La Palma. We used the Intermediate 
Dispersion Spectrograph with the 500mm camera, a 1200\,lines/mm grating and a
TEK charged-coupled device detector to obtain 25 spectra of KPD\,1930+2752
covering 400\AA\ around the H$\alpha$ line with a dispersion of 0.39\AA\ per
pixel. The resolution measured from the full-width at maximum of the arc lines
is 0.9\AA. All the spectra were obtained in a single run of observations of
just over 2 hours on the morning of 17 April 2000. The slit width used was
0.97\,arcsec, which was well matched to seeing estimated from the spatial
profile of the spectra of around 1.3\,arcsec. Observations of a CuNe arc were
obtained before and after the run of observations and every 25 minutes
in-between. The exposure time used for all the spectra was 300s.

 Extraction of the spectra from the images was performed automatically using
optimal extraction to maximize the signal-to-noise of the resulting spectra
(Horne 1986). The arcs associated
with each stellar spectrum were extracted using the same weighting determined
for the stellar image to avoid possible systematic errors due to tilted
spectra.  The wavelength scale was determined from a fourth-order polynomial
fit to measured arc-line positions. The standard deviation of the fit to the 8
spectra lines was typically 0.09\AA. The wavelength scale for an individual
spectrum was determined by interpolation to the time of mid-exposure from the
fits to arcs taken before and after the spectrum to account for the small
amount of drift in the wavelength scale ($<0.1$\AA) due to flexure of the
instrument. Statistical errors on every data point calculated from
photon statistics are rigorously propagated through every stage of the data
reduction.

\section{The spectroscopic orbit}
\begin{figure}
\caption{\label{RVFig} Measured radial velocities for KPD\,1930+2752 from the
H$\alpha$ line (circles) and the HeI\,6678\AA\ (squares) line. Also shown is the
sinusoidal fit (dashed line) used to find good estimates of the amplitude and
systemic velocity for the simultaneous fit to all the spectra.}
\psfig{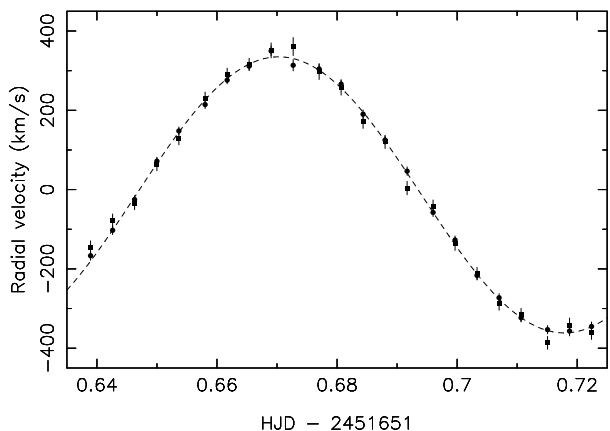}
\end{figure}

 Visual inspection of the H$\alpha$ and HeI\,6678\AA\ spectral lines shows
clearly the Doppler motion expected from a binary star with an orbital period
of $2^{\rm h} 17^{\rm m}$ with a semi-amplitude of $\sim$350\,km\,s$^{-1}$,
confirming the interpretation of Bill\`{e}res et~al. To obtain a more accurate
value of for the semi-amplitude of the orbit, we created a model of the
spectrum composed of three Gaussian profiles for the H$\alpha$ line and a
single Gaussian profile for the HeI\,6678\AA\ line. We used a least-squares
fit to the first spectrum to obtain an initial estimate of the shape of the
model spectrum. We then fixed the shape of the model spectrum and varied only
the position of the lines in a least-squares fit to each of the spectra to
obtain an initial estimate of the radial velocity from each spectrum. These
radial velocities are shown in Fig.~\ref{RVFig}. The spectra were normalised
prior to fitting using a quadratic fit to the continuum either side of the
H$\alpha$ and HeI\,6678\AA\ spectral lines. Only data in the range
6500--6675\AA\ were included in the fit. 

 We then used a simultaneous least-squares fit to all the spectra to
determine the best profile shape and to determine the spectroscopic orbit
simultaneously. The position of the model profile varies from
spectrum-to-spectrum according to the radial velocity predicted by $\gamma +
 K\sin((T-T_0)/P)$, where $\gamma$ is the systemic velocity, $K$ is the
projected orbital speed, $T$ is the time of mid-exposure of the spectrum and
$P$ is the orbital period.  We fixed the value of $P$ at the value given by
Bill\`{e}res et~al. (8217.8s = 0.095111d) since this value is much more
accurately determined by their photometry than can be done from our
spectroscopy. The time $T_0$ corresponds to the point in the orbit  when the
sdB star is closest to the observer.  The smearing of the spectra due to the
motion during the exposure is incorporated into the model. We used independent
values of $\gamma$ for the  H$\alpha$ and HeI\,6678\AA\ spectral lines to allow
for any difference due to pressure shifts. There are many free parameters in
this fit so we first varied the parameters in the spectroscopic orbit while
holding the model profile fixed and then {\it vice versa} so all the
parameters were near their optimum values. We then varied all the parameters
to find the solution given in Table~\ref{RVFitTable}. The fit to the data is
very good judged both by the value of $\chi^2$ and visual inspection of the
residuals. 

\begin{table}
\caption{Results of the simultaneous fit to all the spectra for the model
spectrum and the spectroscopic orbit. Note that the model profile
is convolved with a Gaussian of width 0.9\AA\ prior to fitting to account for
the instrumental resolution.  Parameters shown in bold type are fixed
quantities in the fitting process.\label{RVFitTable}}
\begin{tabular}{lr}
$\gamma$ (H$\alpha$) (km\,s$^{-1}$) & -13.9 $\pm$ 2.2 \\
$\gamma$ (HeI\,6678\AA) (km\,s$^{-1}$) & -12.2 $\pm$ 3.7 \\
$K$ (km\,s$^{-1}$) &  349.3 $\pm$ 2.7 \\
HJD($T_0$)     & 2451651.6466 $\pm$ 0.0001 \\            
$P$ (days)  & {\bf 0.095111}  \\
Gaussian 1 \\
~Rest wavelength (\AA) & {\bf 6562.76} \\
~FWHM(\AA)  & 4.2 $\pm$  0.3 \\
~Depth      & 0.141 $\pm$ 0.007 \\
Gaussian 2 \\
~Rest wavelength (\AA) & {\bf 6562.76} \\
~FWHM(\AA)  & 16.1  $\pm$  0.5 \\
~Depth      & 0.161  $\pm$ 0.006 \\
Gaussian 3 \\
~Rest wavelength (\AA) & {\bf 6562.76} \\
~FWHM(\AA)  & 148  $\pm$ 13  \\
~Depth      & 0.077  $\pm$ 0.013 \\
Gaussian 4 \\
~Rest wavelength (\AA) & {\bf 6678.149} \\
~FWHM(\AA)  & 3.0  $\pm$ 0.2 \\
~Depth      & 0.115  $\pm$ 0.007 \\
No. of data points & 14473 \\
$\chi^2$ & 14067.65 \\
\end{tabular}
\end{table}

\section{Physical properties of the binary}
 
\begin{figure}
\caption{\label{LCFig} The lightcurve of KPD\,1930+2752 after removal of the 
signal due to pulsations from Bill\`{e}res et~al. (2000) with a model
lightcurve (solid line) for the ellipsoidal variability assuming an
inclination of 90$^\circ$. The model lightcurve excluding Doppler boosting is
shown as a dashed line.  }
\psfig{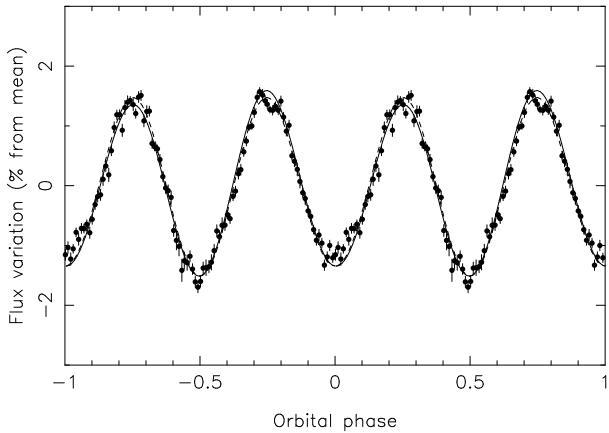}
\end{figure}

 The projected orbital speed, $K=349.3\pm2.7$\,km\,s$^{-1}$, and orbital
period, $2^{\rm h}17^{\rm m}$, immediately imply a minimum mass for the unseen
companion star of 0.42$\pm$0.01\Msolar.  More realistically, the measured
effective temperature and surface gravity of KPD\,1930+2752 (T$_{\rm
eff}$=33\,000K, $\log g = 5.61$) is typical for sdB stars and places it
squarely in the region of the T$_{\rm eff}$\,--\,$\log g$ plane occupied by
models of core helium-burning stars with masses of 0.5\Msolar\  and very thin
hydrogen envelopes ($<0.02\Msolar$, Saffer et~al. 1994), in which
case the  mass of the companion star is at least 0.97$\pm$0.01\Msolar. The
lightcurve of KPD\,1930+2752 observed by Bill\`{e}res et~al. after removal of
the signal due to pulsations is shown in Fig.~\ref{LCFig}. The
quasi-sinusoidal signal with unequal minima characteristic of a star distorted
by a close companion is apparent. We can produce a model lightcurve  using the
physical parameters derived above and assuming that the orbital inclination is
90$^\circ$. The radius of the sdB star implied by the surface gravity and
canonical mass is 0.18$\pm$0.01\Rsolar\ and the separation of the stars is
0.98\Rsolar. The orbit is far too small to contain a normal star of
0.97$\pm$0.01\Msolar\ or more, so we assume that the unseen companion is a
white dwarf star. Other parameters of the model are the gravity darkening
exponent of the sdB star, for which we assume the standard value 
appropriate for radiative stars, and the limb-darkening. The lightcurve was
obtained with a blue-sensitive detector so we use a linear limb-darkening
coefficient of 0.29 which is the mean of the values for U and B filters for a
$\log g=5$, T$_{\rm eff}$=33000 model atmosphere given by Diaz-Cordoves,
Claret \& Gimenez (1995). The precise value of the limb-darkening or gravity
darkening exponent used has very little effect on the lightcurve. An
additional effect included in the model is the Doppler boosting due to the
high orbital velocity. This effect increases the total flux by a factor
$(1-v(t)/\rm c)^3$ where $v$ is the radial velocity at time $t$ and c is
the speed of light. This effect is counteracted by the Doppler shift,
which reduces the effect by a factor $(1-v(t)/\rm c)^2$ for observations on
the Rayleigh-Jeans tail of a black-body spectrum, which is a good
approximation to the spectrum of KPD\,1930+2752 in the optical region. The
overall effect is to make the maxima of the lightcurve asymmetric. This is
shown in Fig.~\ref{LCFig} by plotting a model lightcurves both  with and
without Doppler boosting. The agreement between either model and the observed
lightcurve is excellent. Note that that there has been no attempt made to
optimize the parameters of the model, the lightcurve is prediction based
purely on the measured surface gravity, effective temperature and orbital
velocity together with the assumption that the mass of the sdB star is
0.5\Msolar\ and that the inclination is 90$^\circ$. One effect excluded from
our model is the transit of the companion star across the face of the sdB
star. If the companion is a white dwarf star, its radius will be approximately
0.01\Rsolar\ so the eclipse depth will be about $(0.01/0.18)^2=0.3$\,percent.
There is a hint of just such a feature at the correct phase (0.5) in the
lightcurve, but improved photometry will be required to confirm this feature.

\section{Discussion}

 KPD\,1930+2752 is very similar to the star KPD0422+5421, which is an
sdB--white dwarf binary with an orbital period of 2.16h and an projected
orbital speed of  237$\pm$18\,km\,s$^{-1}$ (Orosz \& Wade 2000). The
ellipsoidal variation in KPD0422+5421 has almost exactly the same amplitude as
that seen KPD\,1930+2752 and the transit of the white dwarf across the face of
the sdB star is seen in high quality lightcurves. 

 Of course, the larger orbital velocity of KPD\,1930+2752 implies a larger
total mass for this binary than KPD0422+5421. The total mass of KPD\,1930+2752
assuming that the sdB star has a mass of 0.5\Msolar\ is at least
$1.47\pm0.01$\Msolar. The significance of this result is that the total mass
exceeds the Chandrasekhar limit for white dwarfs (1.40\Msolar, Hamada \&
Salpeter 1961). KPD\,1930+2752 will merge within about 200 million years due
to a combination of orbital shrinkage through gravitational wave radiation and
the evolutionary expansion of the sdB star. Thus, KPD\,1930+2752 is the first
star to be discovered which is a good candidate for the progenitor of a
Type~Ia supernova due to accretion of helium and other elements heavier than
hydrogen onto the white dwarf which then exceeds the Chandrasekhar mass.

\section{Conclusion}
 We have confirmed the conclusion of Bill\`{e}res et~al. that KPD\,1930+2752
is a binary star in which the sdB star shows ellipsoidal variability and the
orbital period is $2^{\rm h}17^{\rm m}$. The amplitude of the orbital motion
(349.3$\pm$2.7\,km\,s$^{-1}$) combined with the canonical mass for sdB stars
(0.5\Msolar) implies a total mass for the binary of 1.47$\pm$0.01\Msolar. The
unseen companion star is almost certainly a white dwarf star. The binary will
merge within $\sim$200 million years due gravitational wave radiation. The
accretion of helium and other elements heavier than hydrogen onto the white
dwarf which then exceeds the Chandrasekhar mass (1.4\Msolar) is a viable model
for the cause of Type Ia supernovae.  KPD\,1930+2752 is the first star to be
discovered which is a good candidate for the progenitor of a Type~Ia supernova
of this type which may explode on an astrophysically interesting timescale.

\section*{Acknowledgements}
 PFLM was supported by a PPARC post-doctoral grant. The Isaac Newton
Telescope is operated on the island of La Palma by the Isaac Newton Group in
the Spanish Observatorio del Roque de los Muchachos of the Instituto de
Astrofisica de Canarias.

\label{lastpage}
\end{document}